\documentclass[twocolumn]{revtex4-2}
\usepackage{amsmath}
\usepackage{graphicx}
\usepackage{dcolumn}
\usepackage{xcolor}
\usepackage[normalem]{ulem}
\usepackage[section]{placeins}

\usepackage[hyperfootnotes=true]{hyperref}
\hypersetup{colorlinks, citecolor=blue, linkcolor=red, urlcolor=purple}

\begin{document}

\preprint{APS/123-QED}

\title{\textcolor{black}{Aerosol generation by the} splashing of low viscosity drops impacting liquid layers} 

\author{Guillaume Riboux}%
\author{Jos\'e M. Gordillo}
\email{jgordill@us.es}
\affiliation{\'Area de Mec\'anica de Fluidos, Departamento de Ingenier\'ia Aeroespacial y Mec\'anica de Fluidos, Universidad de Sevilla, Avenida de los Descubrimientos s/n 41092, Sevilla, Spain
}%

\date{\today}

\begin{abstract}
 Using theory and numerical simulations, here we describe the early stages of the impact with a velocity $V$ of a drop of radius $R_d$ of a low viscosity liquid such as water against a layer of generic thickness $H$ of the same liquid. Our predictions for the initial velocity $V_t\gg V$ and the diameter $H_t\ll R_d$ of the toroidal rim bordering the edge of the thin lamella which is ejected radially outwards after the impact, are in fair agreement with both the numerical results and also with the experimental measurements reported by Zhang et al [J. Fluid Mech, 690, 5, 2012] \cite{Deegan_2012}. Consequently, the present findings can be employed, for instance, to predict the initial diameters and velocities of the fastest tiny droplets which are ejected right after a rain drop falls on \textcolor{black}{a liquid pool}, an ubiquitous phenomenon with implications in the dispersal of contaminants and in the generation of aerosols \textcolor{black}{and of ice condensation nuclei}.
\end{abstract}
\maketitle

\begin{figure}[!th]
\includegraphics[width=0.35\textwidth]{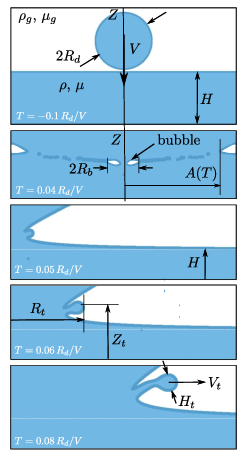}
\caption{\label{fig1} Numerical simulation \textcolor{black}{using \texttt{Basilisk} \citep{Popinet2009,vatsal1}} showing the impact with a velocity $V$ of a drop of radius $R_d$, density $\rho$ and viscosity $\mu$ onto a liquid film of depth $H$ of the same liquid for $We=315$, $Oh=0.0105$, \textcolor{black}{$\rho_g/\rho=10^{-3}$,  $\mu_g/\mu=1.8\times 10^{-2}$} and $h=H/R_d=1.8$.}
\end{figure}

One of the main physical processes involved in the dispersal of contaminants and in the generation of sea spray aerosol is bubble bursting, a phenomenon which can be reliably quantified using very recent results in the literature \cite{PNAS,PRLGanan,JFM2019, JFM2020,PRLBird,JieFengNatPhys,AnnRevDeike,PNASWang, PRF_2023,Abyss,Particulate}. \textcolor{black}{While the production of aerosol particles by raindrop impact on dry soil is also driven by bubble bursting
\cite{RainPorous},
the falling of raindrops on wet forests, puddles, ponds, lakes \cite{AnnRevRain,RainSplash} or the sea \cite{Murphy_2015,Thoroddsen_2024,Marston2026} induce the dispersion of contaminants \cite{RainSplash} and the production of bioaerosols and of ice condensation nuclei \cite{IceNuclei} through a different physical mechanism namely, the ejection and subsequent breakup of the very thin and fast lamella which is issued shortly after the contact between the two liquid masses, see Fig. \ref{fig1}, a physical process known as \emph{drop splashing}}. However, in spite of the number of experimental, theoretical, and numerical studies devoted to analyze the impact of a drop on a liquid layer \cite{THORODDSEN_2002,Josserand_2003,Howison_2005,Deegan_2012,Murphy_2015,Josserand_2016,JosserandRay, Cimpeanu_2018,Sykes_2023,Wang_2023,Thoroddsen_2024,Marston2026}, much less is known about the crucial early stages taking place when raindrops fall \textcolor{black}{on a liquid pool} 
because none of the numerous and recent studies enumerated above addresses the
discrepancies between the very precise experimental measurements reported by \cite{Deegan_2012} over a decade ago using high speed X-ray images and the different quantitative analyses reported so far. \textcolor{black}{Indeed, while the classical theoretical result used to quantify the initial velocity of the lamella ejected after the impact depicted in Fig. \ref{fig1} predicts that \cite{Josserand_2003} $V_{t}\propto \nu^{-1/2} V^{3/2}$, with $\nu$ indicating the liquid kinematic viscosity and $V$ the impact velocity of the drop, the experiments in \cite{Deegan_2012} do not support the prediction in \cite{Josserand_2003} but, instead, suggest a best fit to the data of the type $V_{t}\sim \nu^{-1/4} V^{1.8}$}. 
Motivated by this fact and with the purpose of describing the very early stages of drop impact on a liquid layer we develop a framework similar to the one used to quantify the splashing of drops impacting solid surfaces \cite{PRLSplash}. The results obtained will then be used to predict the initial velocity and the initial radius of curvature of the toroidal rim bordering the ejected liquid sheet, which breaks due to the growth of azimuthal disturbances \cite{THORODDSEN_2002, Deegan_2012,Zhang2012,Agbaglah_2014,JFM2015,Thoroddsen_2018b}, provoking an early splash characterized by the ejection of fast, micron-sized droplets \cite{THORODDSEN_2002,Deegan_2012}, with velocities $V_t\gg V$ and diameters $\sim H_t\ll R_d$ \cite{JFM2015}, see Fig. \ref{fig1}. Our predictions will be validated using the experimental measurements reported by \cite{THORODDSEN_2002,Deegan_2012} and using the results of numerical simulations carried out using \texttt{Basilisk} \citep{Popinet2009,vatsal1}, see the Supplementary Material for details.

\begin{figure*}[!th]
\includegraphics[width=0.8\textwidth]{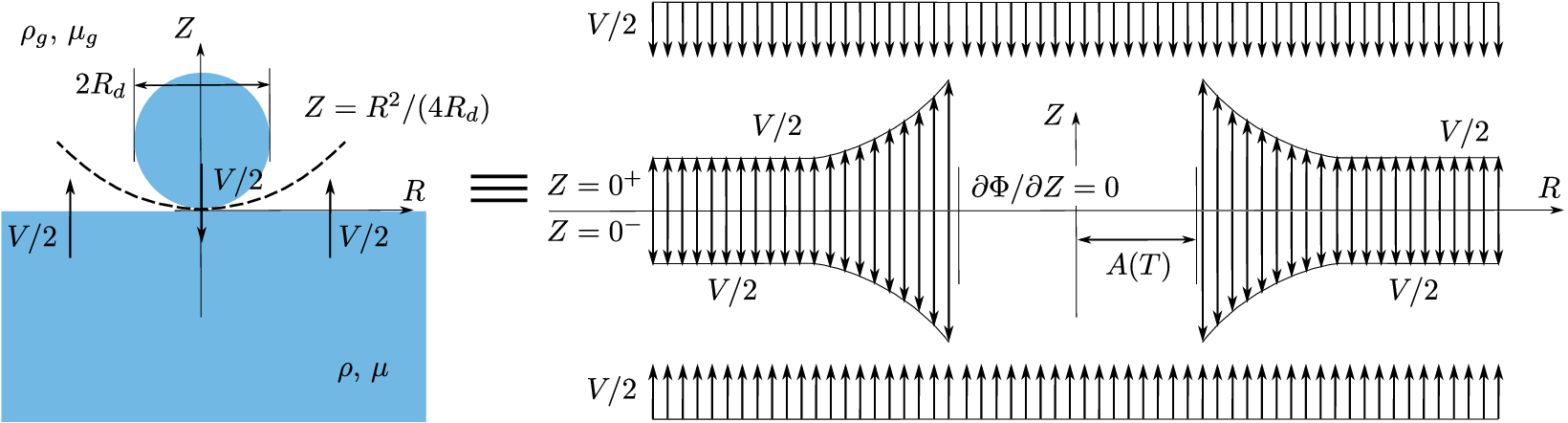}
\caption{\label{fig2} ($a$) Sketch showing the impact of a drop on a liquid layer in a frame of reference moving vertically with the velocity $-V/2$; the dashed line indicates the parabola along which the two liquid masses collide. ($b$) Sketch of the linearized boundary conditions to be satisfied by $\nabla^2\Phi=0$, whose solution provides the vertical velocities given in Eq. (\ref{vfreesurface}), the key ingredient behind the Wagner condition \cite{wagner1932} in Eq. (\ref{wagnercondition}).}
\end{figure*}

We first analyze the case of a liquid pool of thickness $h=H/R_d\rightarrow\infty$ in terms of the dimensionless variables and parameters defined using $R_d$, $V$, $R_d/V$ and $\rho V^2$ as the characteristic values of length, velocity, time and pressure, with $\rho$, $\mu$ and $\sigma$ indicating, respectively, density, viscosity and the interfacial tension coefficient. In the following, lower-case letters will denote dimensionless variables to differentiate them from their dimensional counterparts, written in capitals (e.g. $v_t=V_t/V$, $h_t=H_t/R_d)$ and the subscript $g$ will be used to denote gas quantities\textcolor{black}{: indeed, the numerical results also depend on the ratios $\rho_g/\rho$ and $\mu_g/\mu$, which are kept constant and equal to $10^{-3}$ and $1.8\times 10^{-2}$ unless otherwise specified}.

The type of splashing analyzed here occurs within the \emph{two-jets regime} reported in \cite{Deegan_2012}, which produces the spray reported in \cite{THORODDSEN_2002} for liquids with viscosities equal or similar to that of water. The spray is composed of micron-sized secondary droplets emitted from the toroidal rim bordering the lamella, which destabilizes due to the growth of azimuthal disturbances \cite{THORODDSEN_2002, Deegan_2012,Zhang2012,Agbaglah_2014,JFM2015,Thoroddsen_2018b}. Since splashing \textcolor{black}{and the disintegration of the rim bordering the ejected lamella only} occurs when the values of both the Weber and Reynolds numbers are such that $We=\rho V^2\,R_d/\sigma\textcolor{black}{\gtrsim 250}$ \textcolor{black}{\cite{Deegan_2012}}, $Re=\sqrt{We}\, Oh^{-1}\gg 1$, with $Oh=\mu/\sqrt{\rho R_d\sigma}$, for the cases of interest here, vorticity is confined within very narrow regions at the free interfaces, a fact implying that the dimensionless velocity field is irrotational outside the very thin boundary layers namely, $\mathbf{v}=\nabla\phi$ with $\phi$ indicating the dimensionless velocity potential. Consequently, by virtue of the continuity equation $\nabla\cdot \mathbf{v}=0$, $\phi$ satisfies the Laplace equation $\nabla^2\phi=0$. The potential flow will be described next for $t\ll 1$ in a frame of reference moving with the vertical velocity $-V/2$ (-1/2 in dimensionless terms), with the origin of times $t=0$ set at the instant when the bottom of the drop would contact the liquid layer if the gas was not present. In the moving frame of reference, the free interfaces of both the bottom layer and of the impacting drop -which in the moving frame of reference and for $r\ll 1$ is a parabola of equation $z=r^2/2-t/2$ - are located, neglecting higher order terms \cite{Howison_2005, Cimpeanu_2018}, at $z=0^{\pm}$, with $z=0^+$ indicating the linearized free interface of the drop and $z=0^-$ the corresponding linearized position of the initially flat film. The solution of the Laplace equation subjected to the far field conditions $\partial \phi/\partial z(z\rightarrow \pm \infty)\rightarrow \pm \left(-1/2\right)$, to the impenetrability condition $\partial\phi/\partial z(z=0,r\leq a(t))=0$ and to the boundary condition at the free interfaces $\phi(z=0^\pm,r>a(t))=0$ -deduced using the Euler-Bernoulli equation in the limit $t\ll 1$ \cite{wagner1932,Howison_2005,PRLSplash,Cimpeanu_2018,PRF_2023}, is symmetric with respect to $z=0$ -see the sketch in Fig. \ref{fig2}- and yields the following analytical expression for the vertical velocity at the linearized free interfaces \citep{Lamb,PRLSplash}:
\begin{figure*}[!th]
\includegraphics[width=0.45\textwidth]{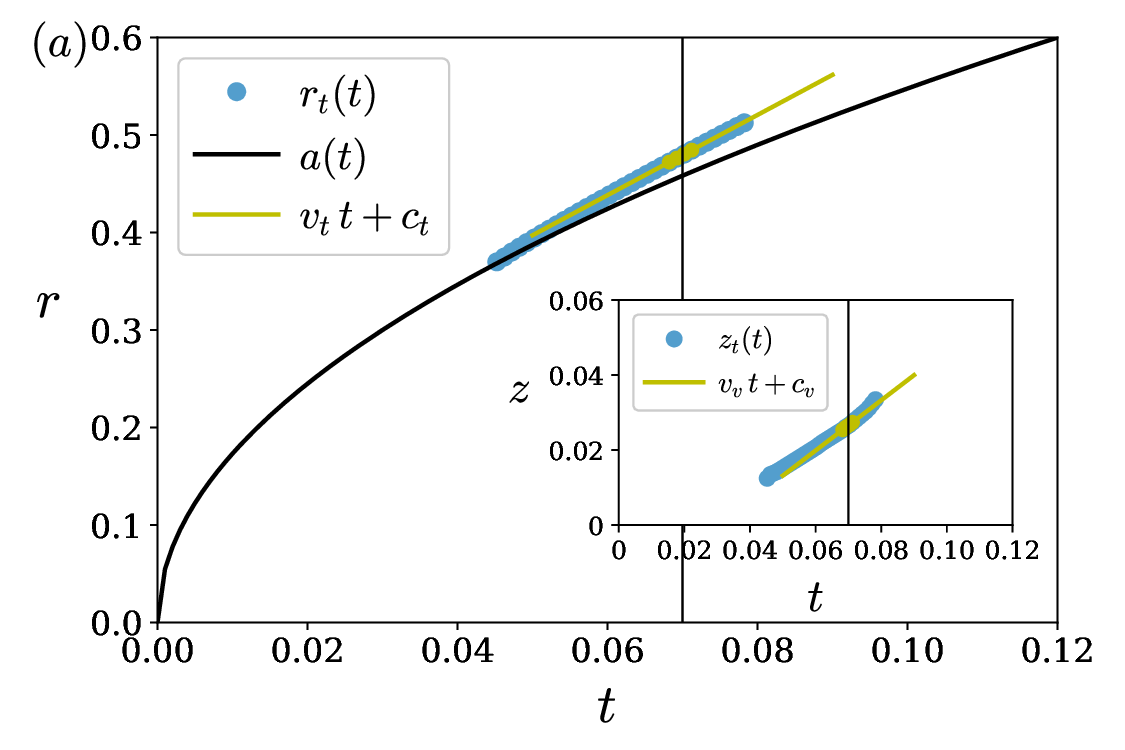}
\includegraphics[width=0.45\textwidth]{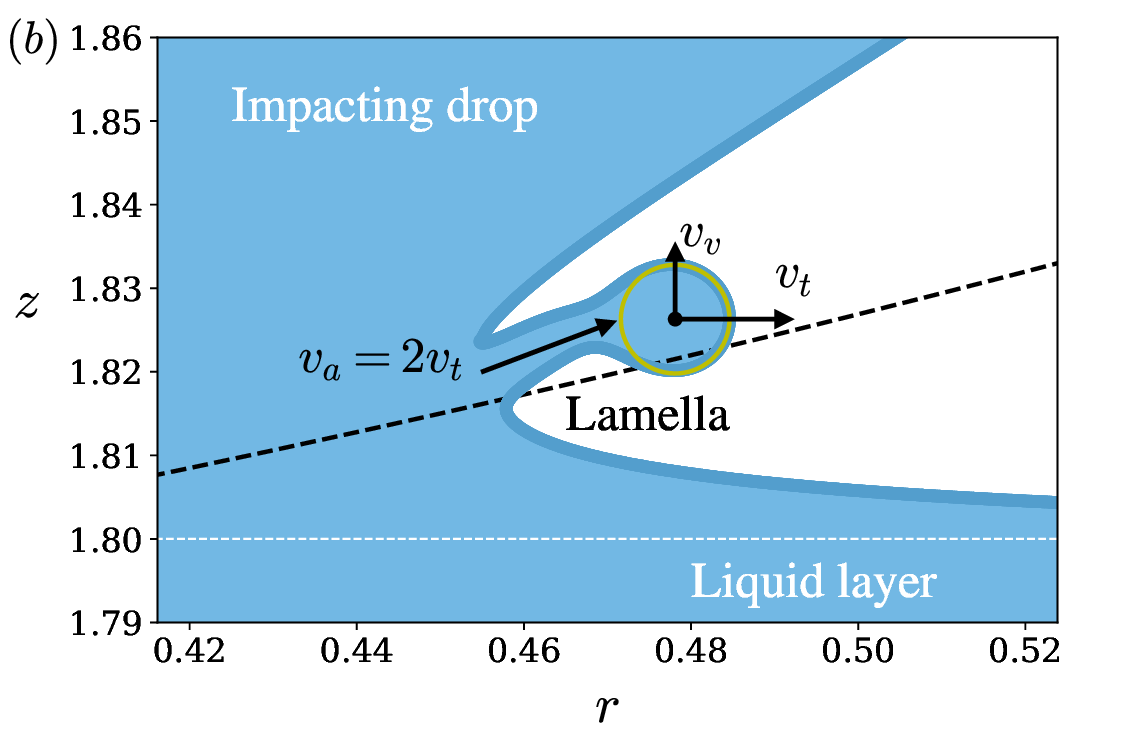}
\caption{\label{fig3}($a$) In blue, \textcolor{black}{numerical values obtained using \texttt{Basilisk}} for the radial, $r_t(t)$, and vertical, $z_t(t)$, positions of the edge of the lamella defined in Fig. \ref{fig1}, \textcolor{black}{see also the Supplementary Material for further details}. The thick black line is \textcolor{black}{the theoretical prediction} $r=a(t)=\sqrt{3t}$ whereas the yellow lines, tangent to $r_t(t)$ and $z_t(t)$, permit to determine the radial and vertical velocities of the edge of the lamella, $v_t$ and $v_v$ respectively, \textcolor{black}{at $t=t_0=0.07$, an instant of time at which the lamella follows a ballistic trajectory -$r_t(t)$ follows a straight line- and, hence, retains its velocity at the ejection instant $t_e<t_0=0.07$ predicted by the solution of Eq. (\ref{eqte}}), see also panel ($b$). ($b$) \textcolor{black}{Numerical} shape of the lamella at $t_0=0.07$ corresponding to the thin vertical black line in figure ($a$). The yellow circle of diameter $h_t$, see also Fig. \ref{fig1}, is the best fit to the local shape of the rim bordering the lamella. The dashed line corresponds to the parabola in Fig. \ref{fig2}a but, in this case, it is represented in the laboratory frame of reference i.e., $z=-t_0/2+r^2/4$. In ($a$)-($b$), $h=1.8$, $We=315$, $Oh=0.0105$.}
\end{figure*}
\begin{equation}
\begin{split}
&\frac{\partial\phi}{\partial z}(r>a(t),z=0^\pm)=\\&=\pm\frac{1}{2}\left(-1-\frac{2}{\pi}\left[\frac{a(t)}{\sqrt{r^2-a^2(t)}}-\arcsin\left(\frac{a(t)}{r}\right)\right]\right)\, .\label{vfreesurface}
\end{split}
\end{equation} 
In Eq. (\ref{vfreesurface}) $a(t)$ indicates the dimensionless radius of the circular region, the wet region in what follows, where the two liquid masses in Fig. \ref{fig2} collide, entrapping a bubble of dimensionless radius $r_b$, see Fig. \ref{fig1}. The radius $a(t)$ is calculated through the so-called Wagner condition \cite{wagner1932}, which is nothing but the time integral of the kinematic boundary condition at the free interfaces and permits to calculate the time taken by a material point located initially at $r=a$, $z=a^2/2$, i.e., a point on the drop interface at $t=0$, to reach the bottom liquid layer. Due to the fact that the vertical velocities at the linearized free interfaces are symmetric with respect to $z=0$, the two liquid masses will collide at the midpoint between their initial positions namely, at $z=r^2/4$ -see Fig. \ref{fig2}- and, therefore, using the expression for the vertical velocity at the upper free interface given in Eq. (\ref{vfreesurface}), the Wagner condition reads \cite{wagner1932}:
\begin{equation}
\begin{split}
&\frac{a^2(t)}{4}-\frac{t}{2}-\\& \frac{1}{\pi}\left(\int_0^t\frac{\textcolor{black}{a(\tau)}\,d\tau}{\sqrt{a^2(t)-\textcolor{black}{a^2(\tau)}}}-\int_0^t\arcsin\left(\frac{\textcolor{black}{a(\tau)}}{a(t)}\right)\,d\tau\right)=0\, ,\label{wagnercondition}
\end{split}
\end{equation}
which is the equation for $a(t)$ deduced and solved in \cite{PRLSplash}, where it was found that $a(t)=\sqrt{3t}$. Consequently, due to the fact that both the local velocity field and the thickness of the lamella depend on $t\ll 1$ through $a(t)$, the results found in \cite{PRLSplash} are also applicable here and, hence, the radial velocity of the wet region is also $\dot{a}(t)=1/2\sqrt{3/t}$; the liquid velocity at the root of the lamella, see Fig. \ref{fig3}, which can be deduced using the Euler-Bernoulli equation along a streamline at the free interface in a frame of reference moving with the radial velocity $\dot{a}$, is also $v_a(t)=2\dot{a}=\sqrt{3/t}$ in the laboratory frame of reference. Note that the solution for the velocity field at the free interface given in Eq. (\ref{vfreesurface}) is not uniformly valid because it diverges for $r\approx a(t)$ where
$\partial \phi/\partial z\approx -(1/\pi)\sqrt{a/2x}$, with $r=a(t)+x$, $x\ll 1$ and, hence, the term $|\nabla\phi|^2/2$ needs to be retained in the Euler-Bernoulli equation in an inner region namely, a lamella of thickness $h_a(t)$, where the blow-up of the outer velocity field in the limit $x\rightarrow 0$ of Eq. (\ref{vfreesurface}), is regularized. The thickness $h_a(t)$ is calculated imposing that the flow rate through the lamella, in the inner region, is the one posed by the outer solution at the root of the lamella namely, $\int_0^{h_a} \sqrt{a/x}\,dx\simeq 2 \dot{a} h_a\Rightarrow h_a\simeq a/\dot{a}^2\propto t^{3/2}$, a classical result due to Wagner \cite{wagner1932,Howison_2005,Cimpeanu_2018}. We then follow the same steps as in \cite{PRLSplash} and, similarly to the case of impact on a solid substrate, we also note from Fig. \ref{fig3} that the edge of the lamella is ejected at a radial velocity $v_t$ that coincides with the wetting velocity at the ejection time $t_e$, i.e. $v_t=\dot{a}(t_e)=1/2\sqrt{3/t_e}$, with $t_e$ deduced from the condition $Dv_t/Dt(t_e)=\ddot{a}(t_e)\propto t_e^{-3/2}$ with $D v_t/Dt<0$ denoting the dimensionless acceleration of the edge of the lamella. The deceleration is caused by the combined effect of both viscous and capillary stresses and is calculated by means of the momentum equation projected in the radial direction, $D v_t/Dt\approx Re^{-1} v_t/h_t^2+We^{-1}/h_t^2$ \cite{PRLSplash}, with $h_t(t_e)\propto h_a(t_e)\propto t_e^{3/2}$ the diameter of the edge of the lamella depicted in Figs. \ref{fig1} and \ref{fig3}. Consequently, the equation for the ejection time $t_e$ is identical to the one deduced in \cite{PRLSplash}:
\begin{equation}
\sqrt{3}/{2}\,Re^{-1}\,t^{-1/2}_e+We^{-1}=c^2t^{3/2}_e\, ,\label{eqte}
\end{equation}
with $c$ a constant  coming from the factor in $h_t\propto h_a$ that was set to $c=1.1$ in \cite{PRLSplash} for the case of drops impacting solid substrates; here, $c=c(h)$ and, for simplicity, here we set $c(h)=1$ in Eq. (\ref{eqte}) and will absorb the dependence of $t_e$ on $c(h)$ by adjusting a factor of order unity, as will be discussed below; notice that, for the reasons in \cite{Hicks_2010} \footnote{\textcolor{black}{For $Oh\ll 1$, $t_e\simeq We^{-2/3}$; moreover, i) the radius from which the lamella is first ejected is $a(t_e)=\sqrt{3\,t_e}\simeq \sqrt{3}\, We^{-1/3}$, ii) the radius of the entrapped bubble is
$r_b\approx 6\,\left(\sqrt{We} \,Oh^{-1}\,\mu/\mu_g\right)^{-1/3}$ \cite{Hicks_2010}. Hence, the larger is the value of the ratio $K=r_b/a(t_e)\simeq 4\,We^{1/6}\, Oh^{1/3} \left(\mu_g/\mu\right)^{1/3}$, the larger is the influence of the entrapped bubble on the ejection of the lamella; in all cases considered here, $K=r_b/a(t_e)<1$}}, the entrapped bubble does not appreciably influence the value of $t_e$ calculated through Eq. (\ref{eqte}) which, in the limit $Oh\ll 1$ of interest here yields $t_e\simeq We^{-2/3}$. \textcolor{black}{All the numerical values of $v_t$, $v_v$ and $h_t$ in Figs. \ref{fig4}-\ref{fig5} are calculated at $t=t_0=0.07$, see Fig. \ref{fig3} because, at this instant of time, the lamella follows a ballistic trajectory \cite{PRLSplash,JFM2015} that retains its initial velocity at $t_e\approx We^{-2/3}<0.07$ when $We>100$, see also} \footnote{\textcolor{black}{For $0.1\leq h\leq 0.5$ and $We\geq 630$, the value of $t_0>t_e$ is just the instant before the rim collides with the falling droplet, $t_0\approx 0.05$}}.

\begin{figure}[!th]
\includegraphics[width=0.48\textwidth]{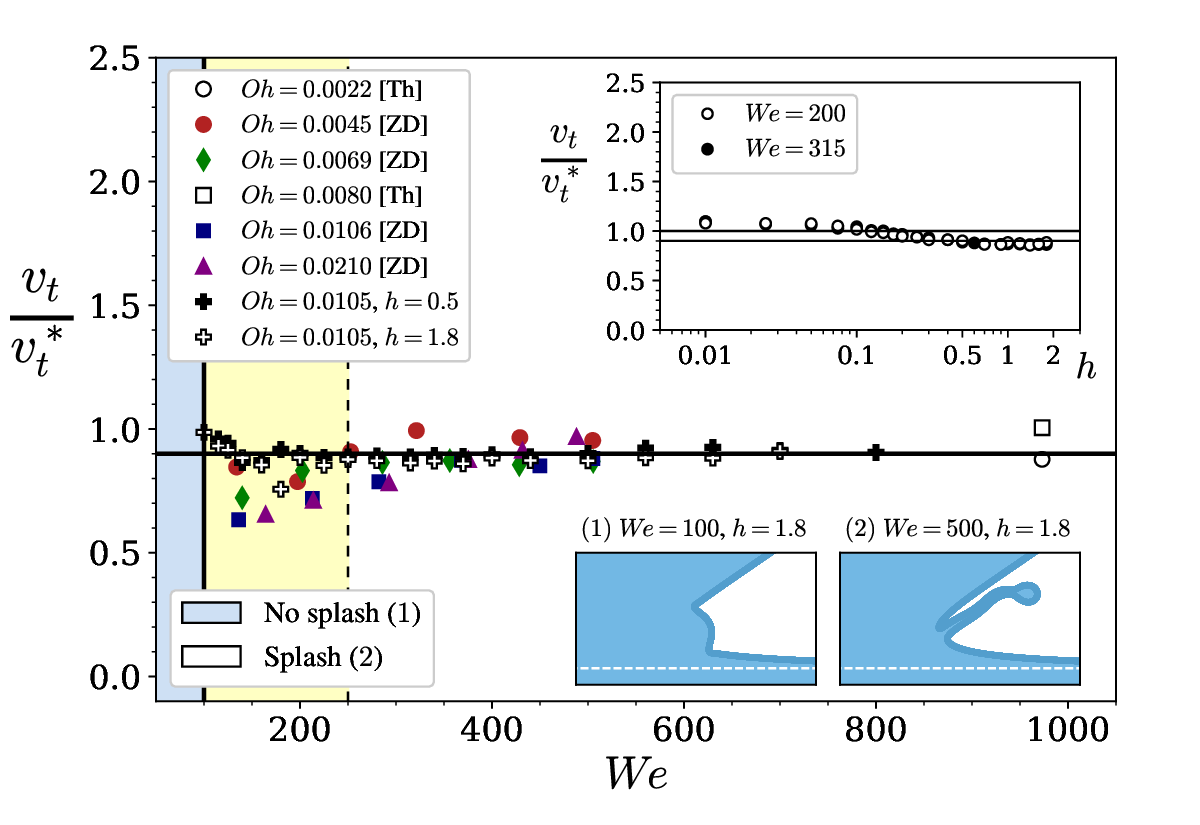}
\caption{\label{fig4} \textcolor{black}{Main:} $v_t/v^*_t$ \textcolor{black}{with $v_t$ the experimental values in Ref. \cite{Deegan_2012} [ZD] and Ref. \cite{THORODDSEN_2002} [Th]} for wide ranges of values $Oh\lesssim 0.02$; $v_t^*=\sqrt{3/t_e}/2$ with $t_e$ the ejection time calculated solving Eq. (\ref{eqte}) -$v_t^*=\sqrt{3/t_e}$ for the experimental data points in \cite{THORODDSEN_2002} \textcolor{black}{[Th]} \textcolor{black}{because the data in this case represent drop velocities: indeed, using the experiments in \cite{Thoroddsen2012}, it was shown in \cite{JFM2015} that the secondary drops emitted directly from the root of the lamella are ejected with the velocity $v_a=\sqrt{3/t}$}. \textcolor{black}{The values of $v_t$ represented using $+$, indicate numerical results; see the Supplementary Material for additional numerical data}. \textcolor{black}{Inset: \emph{numerical} values of $v_t$ for $We=200, 315$ and $Oh=0.0105$ as a function of $h$: notice that $0.9\leq v_t/v^*_t\leq 1$ for over two decades in $h$.} Due to the fact that \textcolor{black}{splash does not occur for $We<100$} \cite{Deegan_2012,Sykes_2023}, $We\geq 100$ in the figure.}
\end{figure}

\begin{figure}[!th]
\includegraphics[width=0.48\textwidth]{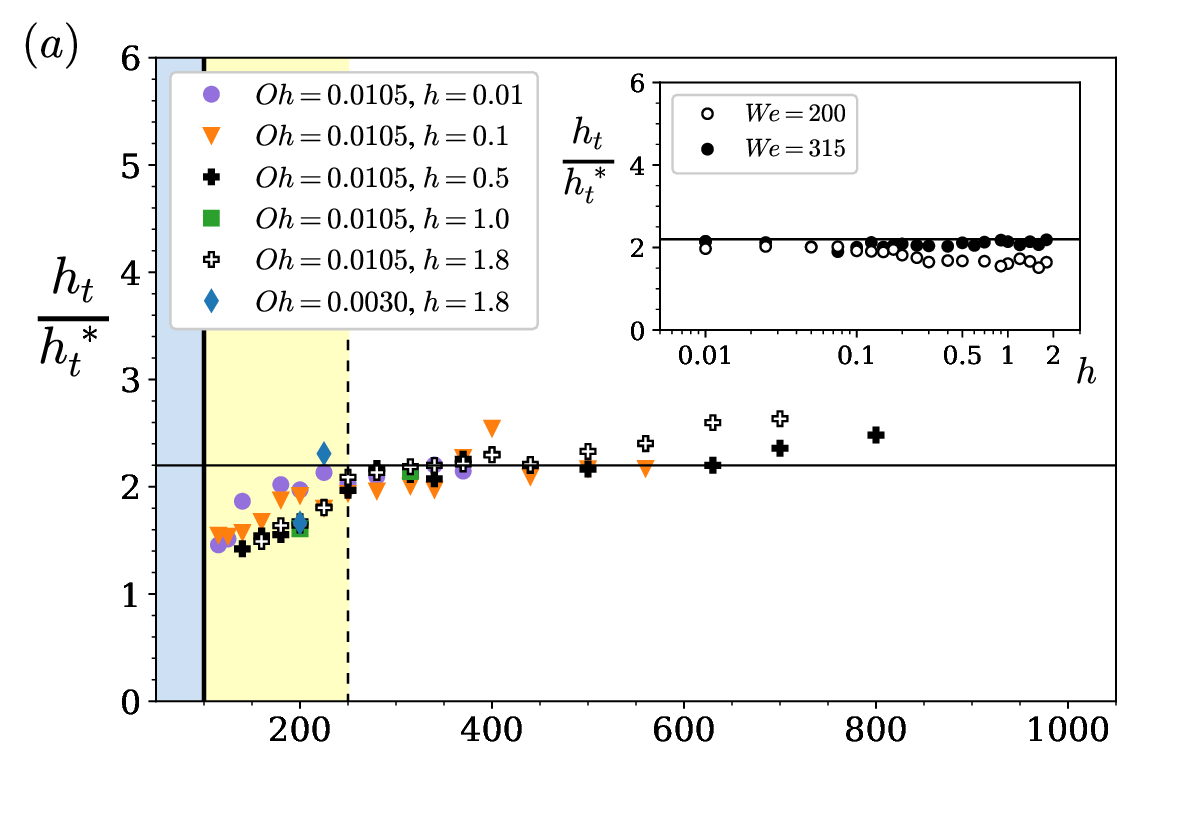}\vspace*{-6mm}
\includegraphics[width=0.48\textwidth]{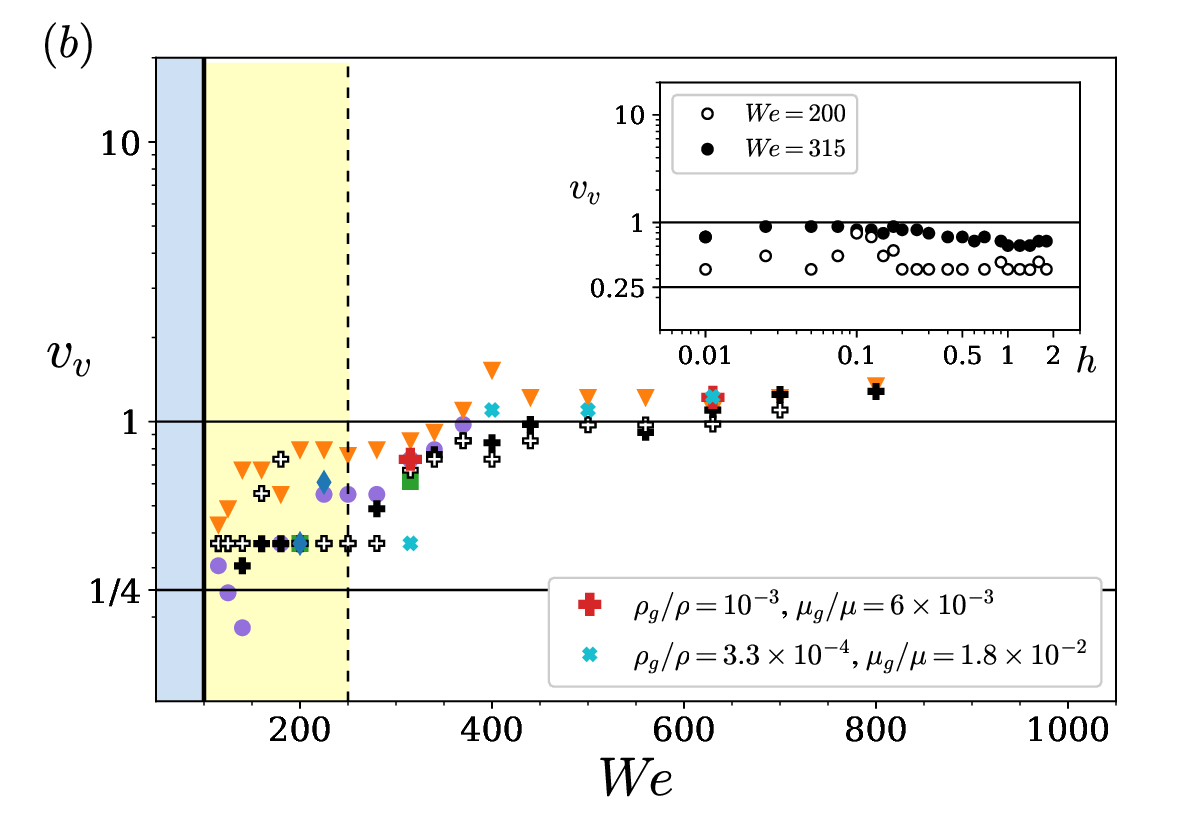}
\caption{\label{fig5} \textcolor{black}{Numerical values of} $h_t/h^*_t$ ($a$) and  $v_v$ ($b$), with $h_t^*=t_e^{3/2}$ and $t_e$ given by Eq. (3). \textcolor{black}{The numerical results in $(b)$ corresponding to $\rho_g/\rho=10^{-3}$, $\mu_g/\mu=0.6\times 10^{-2}$ and to $\rho_g/\rho=0.33\times 10^{-3}$, $\mu_g/\mu=1.8\times 10^{-2}$ have been calculated for $Oh=0.0105$ and $h=1.8$. The insets show the dependence on $h$ of: ($a$) $h_t/h^*_t$, ($b$) $v_v$.}}
\end{figure}
Figure \ref{fig4} compares the experiments of \cite{Deegan_2012,THORODDSEN_2002} and our numerical results with the prediction $v^*_t=1/2\sqrt{3/t_e}$, with $t_e$ given by the solution of Eq. (\ref{eqte}), for wide ranges of values of $We>100$, $Oh$ and $h$ finding \textcolor{black}{that, for the conditions under which the rim disintegrates generating a spray namely, when $We\gtrsim 250$ \cite{Deegan_2012},} $v_t\approx 0.9\times 1/2\sqrt{3/t_e}$ \textcolor{black}{with relative errors $\sim\pm 10\%$}. The results in Fig. \ref{fig4} also reveal that the dependence on $h$ of the initial radial velocity of the ejecta sheet is very small, and the reason for this lies in the fact that, except for $c(h)\approx 1$ in Eq. (\ref{eqte}), the expressions for $t_e$ and $a(t)$ in the limit $h\gg 1$ are identical to those found for the case of drops impacting a rigid substrate considered in \cite{PRLSplash}, a case that can be approached in the limit $h\ll 1$. \textcolor{black}{The comparison between the \emph{numerical} values of} $h_t$ with $h^*_t=t_e^{3/2}$ \textcolor{black}{in Fig. \ref{fig5}a reveals that $h_t\simeq 2.2 t_e^{3/2}$, with relative errors $\sim \pm 10\%$ for over two decades in $h$ when the rim breaks into droplets namely, when $We\gtrsim 250$  \cite{Deegan_2012}; note that the small deviations with respect to our prediction in Fig. \ref{fig5}a are a consequence of the thickening of the rim caused by the fact that relative volume flux into the rim is $\approx h^*_t(v_a-v^*_t)\approx h^*_t\,v^*_t$ \cite{JFM2015} during the time interval $t_0-t_e>0$, with $t_0=0.07$ and $v_a=\sqrt{3/t}$.}

The initial value of the vertical velocity $v_v$ in Fig. \ref{fig3} can also be predicted using Wagner theory \cite{wagner1932} in the limit $h\gg 1$, see also \cite{Cimpeanu_2018}. Indeed, the lamella is ejected at the midpoint between the initial position of the drop and the bottom liquid layer namely, at the parabola of equation $z=r^2/4\Rightarrow dz/dr(r=a(t_e))=a(t_e)/2\ll 1$ -see Fig. \ref{fig2}- and consequently, in the moving frame of reference, $v_v=v_t(t_e) a(t_e)/2=3/4$ which means that, in the laboratory frame of reference, \textcolor{black}{the vertical velocity at the ejection instant is $v_v(t_e)=1/4$}, a value which is much smaller than the tangential velocity, $v_v\ll v_t$ with $v_t\propto t_e^{-1/2}\propto We^{1/3}\gg 1$ because, in the limit $Oh\ll 1$, $t_e\propto We^{-2/3}$ see Eq. (\ref{eqte}). Figure \ref{fig5}b shows that $1/4\lesssim v_v\lesssim 1$ and, \textcolor{black}{in order to check if the deviations of $v_v$ with respect to 1/4 are associated with gas effects \cite{Slingshot,Zhang2012,Thoroddsen_2024}, we have performed simulations at reduced values of $\rho_g/\rho$ and $\mu_g/\mu$, finding only slight differences in the values of $v_v$ but not on $v_t$, see Fig. \ref{fig5}b as well as the Supplementary Material. Then, the reason why $1/4\lesssim v_v\lesssim 1$ in Fig. \ref{fig5}b is because the prediction $v_v=1/4$ refers to the \emph{initial} value at $t=t_e$, whereas the values depicted in Fig. \ref{fig5}b are measured at $t_0>t_e$. Indeed, for instants of time $t>t_e$, the radial velocity at the root of the lamella is $v_a=\sqrt{3/t}$ and, consequently, the vertical velocity at $r=\sqrt{3t}$ in the moving frame of reference is $v_v(t)=v_a(t) a(t)/2=3/2$, which means that $v_v(t)=1$ in the laboratory frame of reference. Then, the results in Fig. \ref{fig5}b are a consequence of the relative flux of vertical momentum entering into the rim, which is accelerated vertically from $v_v(t_e)=1/4$ to $v_v=1$.}

To conclude, here we have made use of Wagner's theory \cite{wagner1932} and its extension accounting for capillary and viscosity effects \cite{PRLSplash}, with the purpose of quantifying the early stages of a drop impacting a liquid layer of arbitrary thickness of a low viscosity liquid such as water. Our main finding here is that, since the diameters $d_t$ of the droplets ejected from the rim bordering the thin ejecta sheet are proportional to the diameter $h_t$ \cite{Agbaglah2013,Agbaglah_2014,JFM2015} and the radial and vertical initial velocities of such droplets, $v_t$ and $v_v$ respectively, are the ones of the toroidal rim \cite{JFM2015}, we find that, in the usual limit $Oh\ll 1$,  $d_t \propto 2\,t^{3/2}_e$, $v_t\propto \sqrt{3/t_e}$, $0.25\leq v_v\lesssim 1$, with $t_e$ the dimensionless ejection time given by Eq. (\ref{eqte}), which depends very weakly on the film thickness $h$ and recovers the analogous equation deduced in \cite{PRLSplash} in the limit $h\ll 1$. Our quantitative description remains valid within the range of dimensionless parameters characterizing the fall of raindrops on a liquid layer of generic depth: \textcolor{black}{for instance, for a rain drop of radius $R_d=3$ mm, $V\approx \sqrt{8g\,R_d\left(\rho/\rho_g\right)/(3\,C_D)}\approx$ 8 m$s^{-1}$, where we assumed that the drag coefficient of the falling drop is $C_D\approx 1$, a spray composed of droplets with diameters $D_t\approx 2 R_d t_e^{3/2}\approx 5$ $\mu$m is ejected initially into the atmosphere with a vertical velocity $V_v\approx 8$ m$s^{-1}$ and a tangential velocity $V_t\approx \sqrt{3} V\,t_e^{-1/2}\approx 150$ m$s^{-1}$ (see Fig. \ref{fig4} and \cite{THORODDSEN_2002}), generating an aerosol containing particles with diameters $\lesssim 5$ $\mu$m. At later times, $t_e<t\ll 1$, particles with diameters $D_p\gtrsim 5$ $\mu$m will also be sprayed: indeed, since the radius of the wetted area increases in time as $R_d\sqrt{3t}$, the rim thickness grows in time as $\approx 2 R_d t^{3/2}$ and the liquid velocity flowing into the lamella decays in time as $V\sqrt{3/t}$, the flux of particles of diameter $D_p$ which will be issued at the instant $t$ into the surrounding atmosphere is approximately given by $\approx 6\pi R^2_d V\,t^{3/2}\left(\overline{C}_+(D_p)+\overline{C}_{-}(D_p)\right)$ with $\overline{C}_\pm(D_p)$ defined in} \footnote{\textcolor{black}{$\overline{C}_\pm(D_p\gtrsim R_d t^{3/2})=0$ and $\overline{C}_\pm(D_p\lesssim R_d t^{3/2})$ refers to the averaged value of the volume concentration of particles of diameter $D_p$ existing at $t=0$ at the interfaces of both the drop ($+$) and the liquid layer ($-$): $\overline{C}_+=1/(2\pi R_d t^{3/2})\int_0^{2\pi}\int_0^{R_d t^{3/2}} C_+(\theta,z,R=R_d\sqrt{3t}) d\theta\,dz$, with $C_{\pm}$ the concentration field at $t=0$: indeed, $R=R_d\sqrt{3t}$ is the radial position where the liquid feeding the lamella comes from \cite{PRLSplash}. $\overline{C}_{-}$ is defined analogously.}}; \textcolor{black}{all the micron-sized solid particles emitted by drop splashing will be later on dispersed by wind}. Hence, our results are applicable to describe the aerosol generated by the early splash of rain drops falling over liquid layers of arbitrary depth, an ubiquitous process with implications in climate \cite{IceNuclei} and in the dispersal of contaminants \cite{AnnRevRain,RainSplash}.

\emph{Acknowledgements} - This work has been supported by the Grant PID2024-156545NB-I00, financed by the Spanish MICIU/AEI/10.13039/501100011033 and by ERDF/EU.

\bibliography{Gordillo_bibv2}

\end{document}